\documentclass[twocolumn]{jpsj3}
\usepackage{txfonts}
\def\runauthor{D. Aoki \textit{et al.}}

\title{Ferromagnetic Quantum Critical Endpoint in UCoAl}

\author{
Dai~{\sc Aoki}$^1$\thanks{E-mail address: dai.aoki@cea.fr}, 
Tristan~{\sc Combier}$^1$, 
Valentin~{\sc Taufour}$^1$,
Tatsuma~D.~{\sc Matsuda}$^{1,2}$, 
Georg~{\sc Knebel}$^1$,
Hisashi~{\sc Kotegawa}$^{1,3}$,
and
Jacques~{\sc Flouquet}$^1$%\\
}

\inst{%
$^1$INAC, SPSMS, CEA Grenoble, 38054 Grenoble, France\\
$^2$ASRC, JAEA, Tokai, Ibaraki 319-1195, Japan\\ 
$^3$Department of Physics, Kobe University, Kobe 657-8501, Japan 
}

\abst{
Resistivity and magnetostriction measurements were performed at high magnetic fields and under pressure on UCoAl.
At ambient pressure, the 1st order metamagnetic transition at $H_{\rm m}\sim 0.7\,{\rm T}$ from the paramagnetic ground state to the field-induced ferromagnetic state 
changes to a crossover at finite temperature $T_{\rm 0}\sim 11\,{\rm K}$.
With increasing pressure, $H_{\rm m}$ linearly increases, 
while $T_0$ decreases and is suppressed at the quantum critical endpoint (QCEP, $P_{\rm QCEP}\sim 1.5\,{\rm GPa}$, $H_{\rm m}\sim 7\,{\rm T}$). 
At higher pressure, the value of $H_{\rm m}$ identified as a crossover continuously increases,
while a new anomaly appears above $P_{\rm QCEP}$ at higher field $H^\ast$ in resistivity measurements.
The field dependence of the effective mass ($m^\ast$) obtained by resistivity and specific heat measurements 
exhibits a step-like drop at $H_{\rm m}$ at ambient pressure.
With increasing pressure, it gradually changes into a peak structure
and a sharp enhancement of $m^\ast$ is observed near the QCEP.
Above $P_{\rm QCEP}$, the enhancement of $m^\ast$ is reduced, and a broad plateau is found between $H_{\rm m}$ and $H^\ast$.
We compare our results on UCoAl with those of the ferromagnetic superconductor UGe$_2$ and the itinerant metamagnetic ruthenate Sr$_3$Ru$_2$O$_7$. 
}

\kword{quantum critical endpoint, effective mass, metamagnetism, ferromagnetism, UCoAl, UGe$_2$}

\begin{document}
\maketitle

\section{Introduction}
Metamagnetism and quantum criticality in strongly correlated electron systems have attracted much interest
as new quantum phases are expected. 
They can be related to unconventional superconductivity, non-Fermi liquid behavior and nematic phases.
In the case of antiferromagnetic (or nearly antiferromagnetic) compounds,
the metamagnetic transition in heavy fermion systems is well documented.~\cite{Flo06_review,Flo10}
A prototype material is CeRu$_2$Si$_2$ where a pseudo-metamagnetic transition occurs at $H_{\rm m}\sim 7.8\,{\rm T}$
and the quantum critical endpoint (QCEP) at which the 1st order metamagnetism collapses exists at negative pressure,
as seen in Ce$_{0.925}$La$_{0.075}$Ru$_2$Si$_2$.~\cite{Fis91,Aok11_CeRu2Si2}

On the other hand, in itinerant ferromagnets, 
the metamagnetic transition from the paramagnetic ground state to the field-induced ferromagnetic state could occur 
when the system is tuned into the paramagnetic ground state at zero field.
Well-known systems are the itinerant ferromagnets UGe$_2$~\cite{Tau10,Kot11}, ZrZn$_2$~\cite{Uhl04}, and the nearly ferromagnetic compound Sr$_3$Ru$_2$O$_7$~\cite{Gri01}.
In particular, UGe$_2$ is an interesting system, because unconventional superconductivity coexisting with ferromagnetism appears near the critical pressure,~\cite{Sax00}
and the superconducting phase is enhanced by the metamagnetic transition between two ferromagnetic phases, FM1 and FM2.~\cite{She01}
Recently we have investigated the ferromagnetic QCEP of UGe$_2$ by resistivity and Hall effect measurements,
and have concluded that the QCEP exists at $\sim 18\,{\rm T}$ and at $\sim 3.5\,{\rm GPa}$.~\cite{Tau10,Kot11}
However severe experimental conditions, namely the high field, very low temperature and high pressure, prevent us from performing precise experiments
above the QCEP to date.
Therefore it is important to find new systems which can be easily tuned to the QCEP.
In this paper, we demonstrate that UCoAl is an ideal system for this kind of study.

UCoAl crystallizes in a hexagonal structure with ZrNiAl-type (space group: $P\bar{6}2m$, No. 189) without inversion symmetry.
Applying the magnetic field along the $c$-axis, 
the paramagnetic ground state 
becomes a field induced ferromagnetic state through the metamagnetic transition at $H_{\rm m}\sim 0.7\,{\rm T}$,
with an induced magnetic moment $M_0 \sim 0.3\,\mu_{\rm B}$.~\cite{And85}
The magnetization curve is very anisotropic between $H \parallel c$-axis and $H\perp c$-axis,
indicating Ising-like behavior.
By doping with Y as a negative pressure or by applying uniaxial stress,~\cite{And07,Ish03}
ferromagnetism appears, indicating that UCoAl is in proximity to ferromagnetic order.
The critical pressure where the Curie temperature $T_{\rm Curie}$ is suppressed to zero 
will be negative ($P_{\rm c}\sim -0.2\,{\rm GPa}$)~\cite{Mus99}
and the ground state at ambient pressure is already the paramagnetic one.
$H_{\rm m}$ linearly increases with increasing hydrostatic pressure.~\cite{Mus99}
The metamagnetic transition at $H_{\rm m}$ is identified to be of 1st order 
by the clear hysteresis between increasing and decreasing fields.
Applying pressure, this hysteresis starts to be suppressed, 
which may suggest that the 1st order transition will terminate at a QCEP.
However there are no experimental reports which clarify the existence of the QCEP at high pressures above $1.2\,{\rm GPa}$.
Here we report the experimental evidence for the QCEP in UCoAl 
detected by resistivity and magnetostriction measurements at high fields and at high pressures.

%%%%%%%%%%%%%%%%%%%%%%%%%%%%%%%%%%%%%%%%%%%%%%%%%%%%%%%%%%%%%%%%%%%%%%%%%%%%%%%%%
\section{Experimental}
Single crystals of UCoAl were grown using the Czochralski method in a tetra-arc furnace.
The starting materials of U (purity: $99.95\,{\%}$-3N5) , Co (3N) and Al (5N) were melted on a water-cooled copper hearth under a high purity Ar atmosphere gas.
The ingot was turned over and was melted again.
This process was repeated several times in order to obtain a homogeneous polycrystalline ingot.
The ingot was subsequently pulled with a rate of $15\,{\rm mm/hr}$ for single crystal growth.
The obtained single crystal ingot was cut using a spark cutter
and was oriented by X-ray Laue photograph, displaying very sharp Laue spots.
The resistivity using a sample with rectangular shape ($0.5\times 0.5 \times 1\,{\rm mm}^3$, $c$-axis long) was measured by a four probe AC method ($f\sim 17\,{\rm Hz}$) 
for the electrical current along $[10\bar{1}0]$ direction ($J \perp c$-axis) at low temperatures down to $0.1\,{\rm K}$
and at high fields up to $16\,{\rm T}$ for the field along $[0001]$ direction ($H \parallel c$-axis).
The residual resistivity ratio (RRR) was approximately $10$.
The magnetostriction was measured using strain gauges by the active dummy method with a wheatstone bridge at temperatures down to $2\,{\rm K}$ using a lock-in amplifier ($f \sim 17\,{\rm Hz}$).
The gauge was glued on the surface of the sample with the dimension of $2\times 2\times 0.4\,{\rm mm}^3$, 
so that it detects the dilatation along the $c$-axis ($\Delta L_c$).
Both resistivity and magnetostriction measurements were performed under pressure up to $2.4\,{\rm GPa}$ using a CuBe-NiCrAl hybrid type piston cylinder cell with Daphne oil 7373 as a pressure medium.
The pressure was determined by the superconducting transition temperature of Pb.
The sharp transition assures that the pressure gradient is small ($<0.05\,{\rm GPa}$) for all pressure range.
However, the pressure gradient for magnetostriction measurements with a relatively large sample is larger,
which is approximately $0.1\,{\rm GPa}$ in maximum.~\cite{Koy07}
The Daphne oil 7373 solidifies at $2.2\,{\rm GPa}$ at room temperature,~\cite{Yok07} thus
the hydrostaticity is good at least up to $\sim 2\,{\rm GPa}$ at low temperatures.
For comparison with the field dependence of the resistivity $A$ coefficient,
the specific heat was measured by the relaxation method at ambient pressure
under magnetic field up to $9\,{\rm T}$ and at low temperature down to $0.45\,{\rm K}$.
Angular dependences of $H_{\rm m}$ from $H \parallel c$-axis to $H \perp c$-axis ($H\parallel [10\bar{1}0]$) 
were also measured at ambient pressure by magnetization and magnetostriction.
The magnetization was measured at temperatures down to $2\,{\rm K}$ and at magnetic fields up to $5.5\,{\rm T}$ using a SQUID magnetometer.
The magnetostriction was measured employing the same manner as under pressure at temperatures down to $2\,{\rm K}$ and at fields up to $9\,{\rm T}$ using a horizontal-axis sample rotator.

%%%%%%%%%%%%%%%%%%%%%%%%%%%%%%%%%%%%%%%%%%%%%%%%%%%%%%%%%%%%%%%%%%%%%%%%%%%%%%%%%
\section{Experimental results}
\subsection{Ambient pressure and angular dependence}
\begin{figure}[tbh]
\begin{center}
\includegraphics[width=1 \hsize,clip]{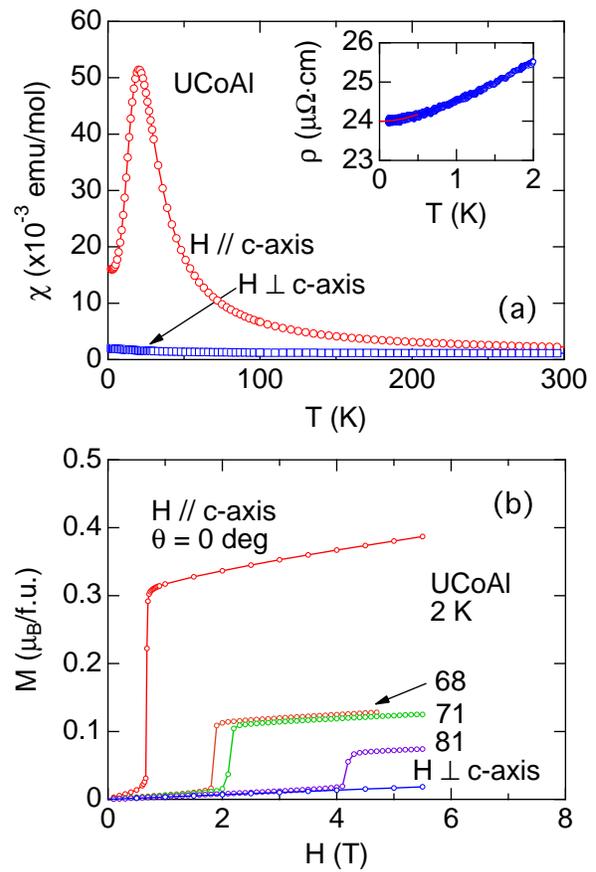}
\end{center}
\caption{(Color online) (a)Temperature dependence of the susceptibility at $1\,{\rm kOe}$ for $H\parallel c$-axis and $H \perp c$-axis in UCoAl. (b)Magnetization curves with increasing field at $2\,{\rm K}$ at different field angles $\theta$ from $H \parallel c$-axis to $H\perp c$-axis. The inset of panel (a) shows the temperature dependence of resistivity at zero field, indicating that $T^2$ dependence is preserved at least up to $0.5\,{\rm K}$ within experimental precision.}
\label{fig:UCoAl_sus_mag}
\end{figure}

Figure~\ref{fig:UCoAl_sus_mag} show the susceptibility and magnetization at $2\,{\rm K}$.
A very anisotropic susceptibility response 
is found between $H \parallel c$ and $H \perp c$-axis,
indicating the Ising-type behavior, which is in good agreement with the previous results.~\cite{Hav97,Mat99,Mus01}
For $H\parallel c$-axis, the susceptibility shows a broad maximum around $20\,{\rm K}$,
while no anomaly is observed for $H\perp c$-axis.
This behavior is typical for heavy fermion systems, such as CeRu$_2$Si$_2$, UPt$_3$, URu$_2$Si$_2$.
Applying the magnetic field along $c$-axis at low temperature,
a sharp metamagnetic 1st order transition is observed at $H_{\rm m}\sim 0.7\,{\rm T}$,
with a hysteresis between the upsweep and downsweep measurements.

First we focus on the angular dependence of the metamagnetic transition,
since in the well-known ferromagnetic systems, 
such as URhGe~\cite{Aok01,Lev07,Miy09,Aok11_ICHE}, UCoGe~\cite{Huy07,Aok09_UCoGe,Iha10} 
and Sr$_3$Ru$_2$O$_7$ (nearly ferromagnetic compound)~\cite{Gri03},
the field angle is found to be a tuning parameter for the quantum singularities, as pressure is.
By increasing the field angle $\theta$ from $H\parallel c$-axis to $H\perp c$-axis,
$H_{\rm m}$ shifts to higher fields proportional to $1/\cos \theta$.
It should be noted that the induced moment just above $H_{\rm m}$ decreases with $\theta$,
which roughly follows $\cos \theta$ dependence.
This is different from the pressure response of magnetization up to $1.2\,{\rm GPa}$ for $H\parallel c$-axis,
where the induced moment just above $H_{\rm m}$ remains almost at the same value, $\sim 0.3\,\mu_{\rm B}$,
while $H_{\rm m}$ is monotonously increased with pressure.~\cite{Mus99}
The field angle response is obviously different from the pressure response in UCoAl.
%========================================================================================
%========================================================================================

\begin{figure}[tbh]
\begin{center}
\includegraphics[width=1 \hsize,clip]{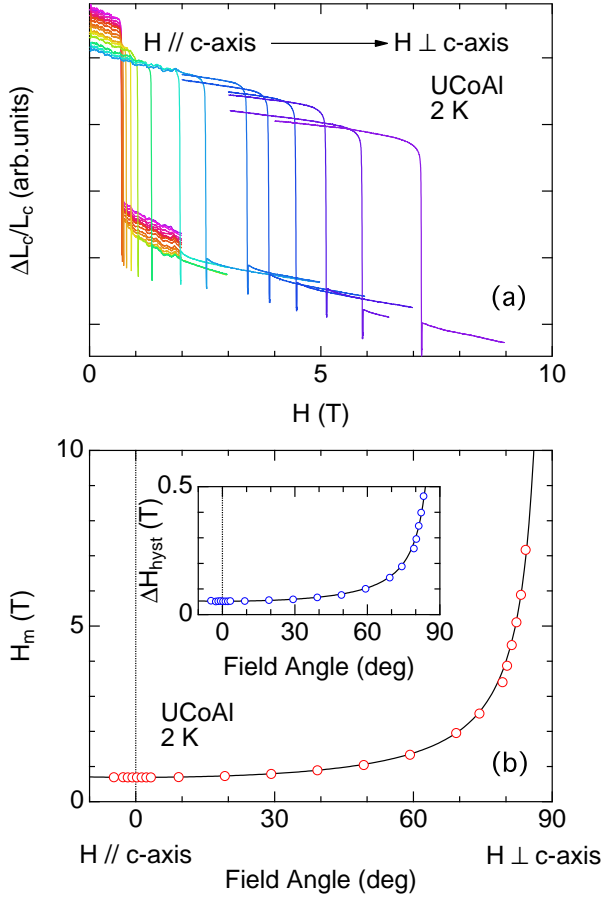}
\end{center}
\caption{(Color online) (a)Field dependences of the magnetostriction with increasing fields at $2\,{\rm K}$ at different field angles from $H\parallel c$-axis to $H\perp c$-axis in UCoAl. (b)Angular dependence of the metamagnetic transition field $H_{\rm m}$ at $2\,{\rm K}$ for upsweep measurements. The inset of panel (b) shows the angular dependence of the hysteresis of $H_{\rm m}$ between upsweep and downsweep. The solid lines correspond to a $1/\cos \theta$ dependence.}
\label{fig:UCoAl_MS_angdep}
\end{figure}

Figure~\ref{fig:UCoAl_MS_angdep}(a) show the field dependence of the magnetostriction $\Delta L_c/L_c$ for different field angles at $2\,{\rm K}$.
Sharp drops of magnetostriction due to the metamagnetic transition in agreement with the previous results~\cite{Hon00} are observed at $H_{\rm m}$,
which is increased with increasing field angle.
As shown in Fig.~\ref{fig:UCoAl_MS_angdep}(b), $H_{\rm m}$ increases following a $1/\cos \theta$ dependence,
at least up to $7.2\,{\rm T}$ at $84\,{\rm deg}$.
The magnitude of the jump of magnetostriction retains a large value even at high field angles.
Furthermore the hysteresis at $H_{\rm m}$, $\Delta H_{\rm hyst}$ also increases with field angle, following the $1/\cos\theta$ dependence,
as shown in the inset of Fig.~\ref{fig:UCoAl_MS_angdep}(b).
It is noted that the width of metamagnetic transition ($\Delta H_{\rm m}\sim 0.02\,{\rm T}$) shows no significant increase with field angle,
indicating that the sharp metamagnetic transition is retained up to $84\,{\rm deg}$.
These results indicate that the 1st order nature is very robust against the field angle, 
which cannot be a tuning parameter to a QCEP in UCoAl at least at ambient pressure.

%========================================================================================
%========================================================================================

It should be noted that Fermi liquid properties are observed in the resistivity measurements, as shown in the inset of Fig.~\ref{fig:UCoAl_sus_mag}(a), where the resistivity follows the $T^2$ dependence below $0.5\,{\rm K}$,
in good agreement with very low temperature specific heat data.~\cite{Mat99}.
Thus, the achievement of a very low temperature ($T< 0.5\,{\rm K}$) is a necessary condition to observe the Fermi liquid regime at ambient pressure.

\subsection{Pressure study}
\begin{figure}[tbh]
\begin{center}
\includegraphics[width=1 \hsize,clip]{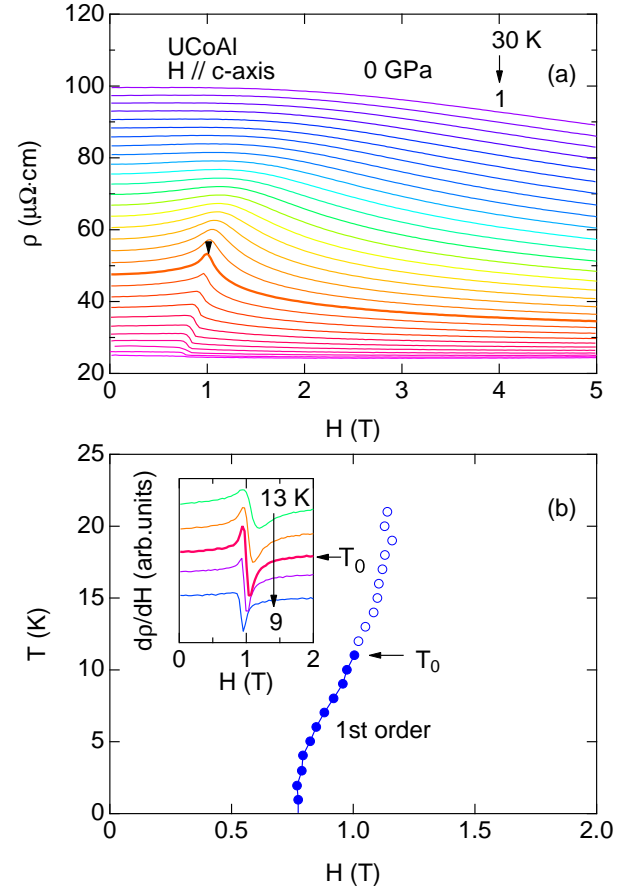}
\end{center}
\caption{(Color online) (a)Magnetoresistance with increasing fields for $H\parallel c$-axis at various constant temperatures from $1$ to $30\,{\rm K}$ with $1\,{\rm K}$ step in UCoAl. (b)Temperature variation of the metamagnetic transition field $H_{\rm m}$.
The inset shows the field derivative of magnetoresistance. 
The thick line corresponds to data at $T_0=11\,{\rm K}$,
at which the metamagnetic transition changes from the 1st order to the crossover.}
\label{fig:UCoAl_resist_ambient}
\end{figure}

Figure.~\ref{fig:UCoAl_resist_ambient}(a) shows the field dependence of the magnetoresistance at ambient pressure
for different constant temperatures.
Anomalies due to the metamagnetic transition from the paramagnetic state to the field-induced ferromagnetic state are clearly observed around $H_{\rm m}\sim 0.7\,{\rm T}$.
The anomaly with a small step-like decrease at low temperatures gradually changes into a sharp peak
around $10\,{\rm K}$.
Further increasing temperature, the anomaly is smeared out.
Figure~\ref{fig:UCoAl_resist_ambient}(b) shows the temperature variation of $H_{\rm m}$.
$H_{\rm m}$ slightly shifts to higher fields with increasing temperature
implying that the ferromagnetic correlations play a main role for the metamagnetism in UCoAl, as observed just above $P_{\rm c}$ in UGe$_2$.~\cite{Tau10}
At low temperatures, $H_{\rm m}$ is obviously identified as the 1st order transition, 
since hysteresis between the upsweep- and the downsweep-field is observed in magnetoresistance.
With increasing temperature, $H_{\rm m}$ changes from  1st order to a crossover at $T_0$.
Here we determined $T_0$ from the field derivative of magnetoresistance, $d\rho/dH$.
As shown in the inset of Fig.~\ref{fig:UCoAl_resist_ambient}(b),
$d\rho/dH$ reveals sharp maximum and minimum below and above $H_{\rm m}$.
When the peak-to-peak amplitude of $d\rho /dH$ becomes maximum, meaning an acute peak of the magnetoresistance,
we define this temperature as $T_0$.
In Fig.~\ref{fig:UCoAl_resist_ambient}, $T_0$ is found to be $11\,{\rm K}$.
This value is in good agreement with that obtained from the hysteresis of $H_{\rm m}$ in the magnetostriction measurements, as mentioned later,
supporting the validity of our definition for $T_0$.
%========================================================================================
%========================================================================================

\begin{figure}[tbh]
\begin{center}
\includegraphics[width=1 \hsize,clip]{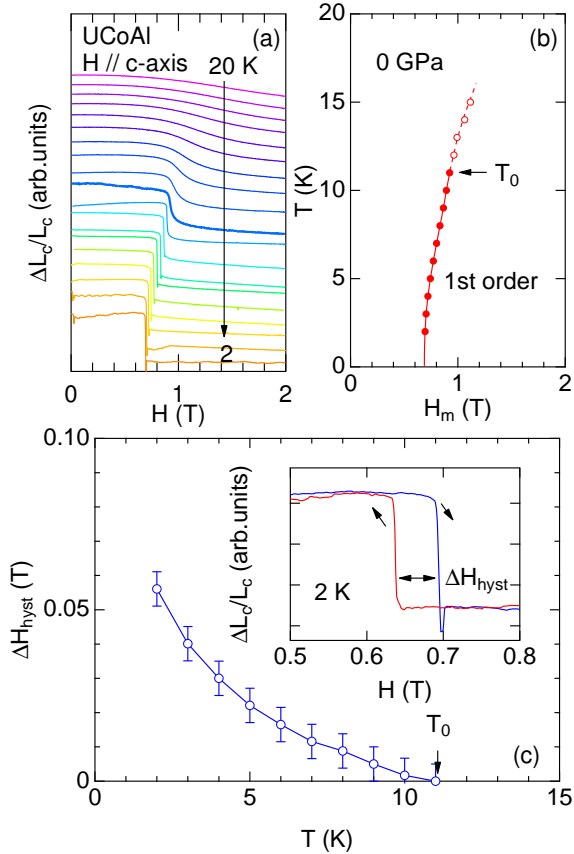}
\end{center}
\caption{(Color online) (a)Field dependence of the magnetostriction along $c$-axis for $H\parallel c$-axis at various constant temperatures from $20$ to $2\,{\rm K}$ with $1\,{\rm K}$. (b)Temperature evolution of the metamagnetic transition field $H_{\rm m}$. The inset of panel(c) displays the hysteresis of $H_{\rm m}$ at $2\,{\rm K}$ between increasing and decreasing field. (c)Temperature dependence of the hysteresis field $\Delta H_{\rm m}$.}
\label{fig:UCoAl_MS_ambient}
\end{figure}

Figure~\ref{fig:UCoAl_MS_ambient}(a) shows the field dependence of the magnetostriction at ambient pressure for different constant temperatures.
Sharp drops of the magnetostriction are observed at $H_{\rm m}$ at low temperatures.
This is in good agreement with the previous report.~\cite{Hon00}
At high temperatures, the anomalies at $H_{\rm m}$ become broad and finally they are smeared out.
As shown in the inset of Fig.\ref{fig:UCoAl_MS_ambient}(c), 
a hysteresis ($\Delta H_{\rm hyst} \sim 0.06\,{\rm T}$) is clearly observed at $2\,{\rm K}$
between up and down field sweeps
indicating the 1st order transition at low temperatures.
$\Delta H_{\rm hyst}$ decreases with increasing temperatures 
and finally becomes zero at $T_0 \sim 11\,{\rm K}$, which is in good agreement with the value 
obtained by the magnetoresistance measurements as shown in Fig.~\ref{fig:UCoAl_resist_ambient}.
The transition at $H_{\rm m}$ changes to a crossover above $T_0$.
%========================================================================================
%========================================================================================

\begin{figure}[tbh]
\begin{center}
\includegraphics[width=1 \hsize,clip]{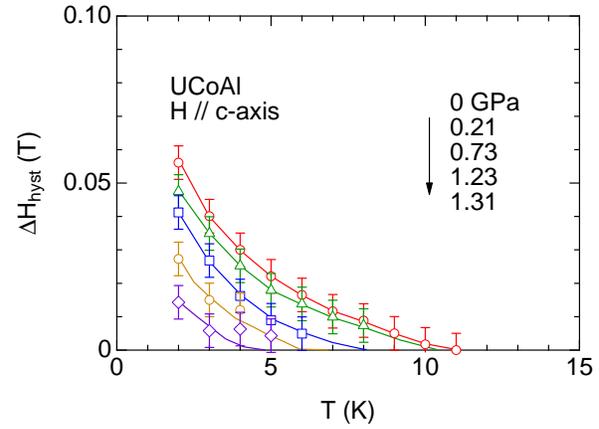}
\end{center}
\caption{(Color online) Temperature dependence of the hysteresis field $\Delta H$ at various pressures in UCoAl.}
\label{fig:UCoAl_MS_Hysteresis}
\end{figure}

As shown in Fig.~\ref{fig:UCoAl_MS_Hysteresis},
the hysteresis is suppressed by applying pressure.
With increasing pressure, $\Delta H_{\rm hyst}$ at $2\,{\rm K}$ is markedly suppressed 
from $0.06\,{\rm T}$ at ambient pressure to $\sim 0.015\,{\rm T}$ at $1.31\,{\rm GPa}$.
No hysteresis was observed at higher pressures ($P \ge 1.75\,{\rm GPa}$).
Correspondingly, $T_0$ decreases with pressures.
It is, however, difficult to evaluate the value of $T_0$ precisely for different pressures,
because $T_0$ is too small for the magnetostriction measurements using strain gauges,
giving rise to relatively large error bars of $\Delta H_{\rm hyst}$.
Nevertheless, the present results imply that the 1st order transition at $H_{\rm m}$ will
terminate at high pressure probably above $P_{\rm QCEP} \sim 1.5\,{\rm GPa}$.
%========================================================================================
%========================================================================================

\begin{figure}[tbh]
\begin{center}
\includegraphics[width=1 \hsize,clip]{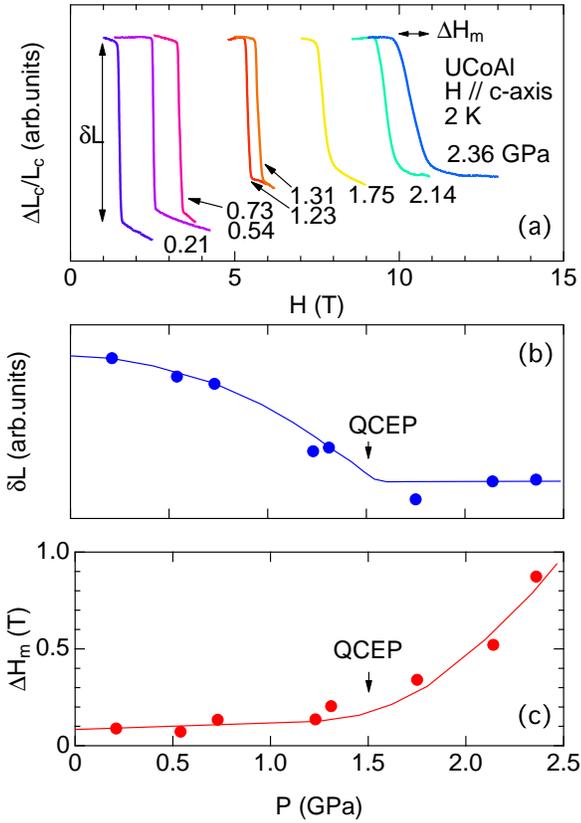}
\end{center}
\caption{(Color online) (a)Field dependence of the magnetostrictions for $H \parallel c$-axis at various pressures at $2\,{\rm K}$ in UCoAl. (b)Pressure dependence of the amplitude of jumps in magnetostriction at $H_{\rm m}$. (c)Pressure dependence of the transition width. QCEP is located around $1.5\,{\rm GPa}$.}
\label{fig:UCoAl_Pdep_jump}
\end{figure}

Figure~\ref{fig:UCoAl_Pdep_jump}(a) shows the field dependence of the magnetostrictions at $2\,{\rm K}$ at various pressures.
Increasing pressure, $H_{\rm m}$ increases linearly to high fields 
and reaches $11\,{\rm T}$ at $2.4\,{\rm GPa}$.
Here we define $H_{\rm m}$ from a midpoint of the magnetostriction jump.
The amplitude of the jump ($\equiv \delta L$) decreases with pressure
and most likely remains constant above $P_{\rm QCEP} \sim 1.5\,{\rm GPa}$, as shown in Fig.~\ref{fig:UCoAl_Pdep_jump}(b).
On the other hand, the transition width ($\equiv \Delta H_{\rm m}$) is almost constant or slightly increases up to $1.5\,{\rm GPa}$, and 
then rapidly increases with further increasing pressure, as shown in Fig.~\ref{fig:UCoAl_Pdep_jump}(c).
One can speculate that the quantum critical endpoint (QCEP) is located around $P_{\rm QCEP}\sim 1.5\,{\rm GPa}$.
At first aprroximatation, the magnetic contribution of the magnetovolume effect $\Delta V_{\rm m}/V_{\rm m}$ is
related to the magnetization $M$ via the relation, $\Delta V_{\rm m}/V_{\rm m} \propto M^2$.
According to the previous magnetization measurements under pressure up to $1.2\,{\rm GPa}$,~\cite{Mus99}
the initial slope of magnetization is unchanged, but $H_{\rm m}$ increases with pressure,
while the induced moment above $H_{\rm m}$ is constant ($\sim 0.3\,\mu_{\rm B}$).
The present results in Fig.~\ref{fig:UCoAl_Pdep_jump} should be related to
the pressure response of magnetization curve.
%========================================================================================
%========================================================================================

\begin{figure}[tbh]
\begin{center}
\includegraphics[width=1 \hsize,clip]{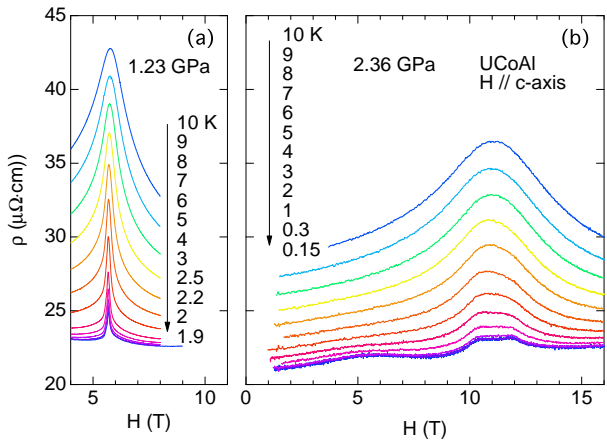}
\end{center}
\caption{(Color online) Field dependence of the magnetoresistance at various fixed temperatures at two different pressures, $1.23\,{\rm GPa}$ ($<P_{\rm QCEP}$) and $2.36\,{\rm GPa}$ ($>P_{\rm QCEP}$) for $H \parallel c$-axis in UCoAl.}
\label{fig:UCoAl_Hdep_resist}
\end{figure}

Figure~\ref{fig:UCoAl_Hdep_resist} shows the magnetoresistance at different constant temperatures at 
two different pressures below and above $P_{\rm QCEP}$.
At $1.23\,{\rm GPa}$ ($< P_{\rm QCEP}$), the magnetoresistance shows the very sharp peak at $H_{\rm m}=5.7\,{\rm T}$
at $1.9\,{\rm K}$, which is quite different from that at ambient pressure
where the step-like behavior is observed at low temperatures.
From the analysis of $d\rho/dH$ as in the inset of Fig.~\ref{fig:UCoAl_resist_ambient}(b),
$T_0$ at $1.23\,{\rm GPa}$ is found to be $4\,{\rm K}$,
which is reduced from the original value, $T_0=11\,{\rm K}$ at ambient pressure.

On the other hand, at $2.36\,{\rm GPa}$ ($> P_{\rm QCEP}$), the magnetoresistance shows a plateau around $H_{\rm m}$
at low temperatures.
Two kinks are observed at $H_{\rm m}=10.5\,{\rm T}$ and $H^\ast = 12\,{\rm T}$ at the lowest temperature.
The former kink at $10.5\,{\rm T}$ is fairly in good agreement with the results of magnetostriction 
as shown in Fig.~\ref{fig:UCoAl_Pdep_jump}(a).
However, the latter kink at $12\,{\rm T}$ was only observed in the magnetoresistance measurements,
while no anomaly was detected at $12\,{\rm T}$ in the magnetostriction measurements.
Two kinks are merged and broaden at high temperatures.
Interestingly, another broad hump is observed around $5\,{\rm T}$ only at low temperatures.
%========================================================================================
%========================================================================================

\begin{figure}[tbh]
\begin{center}
\includegraphics[width=1 \hsize,clip]{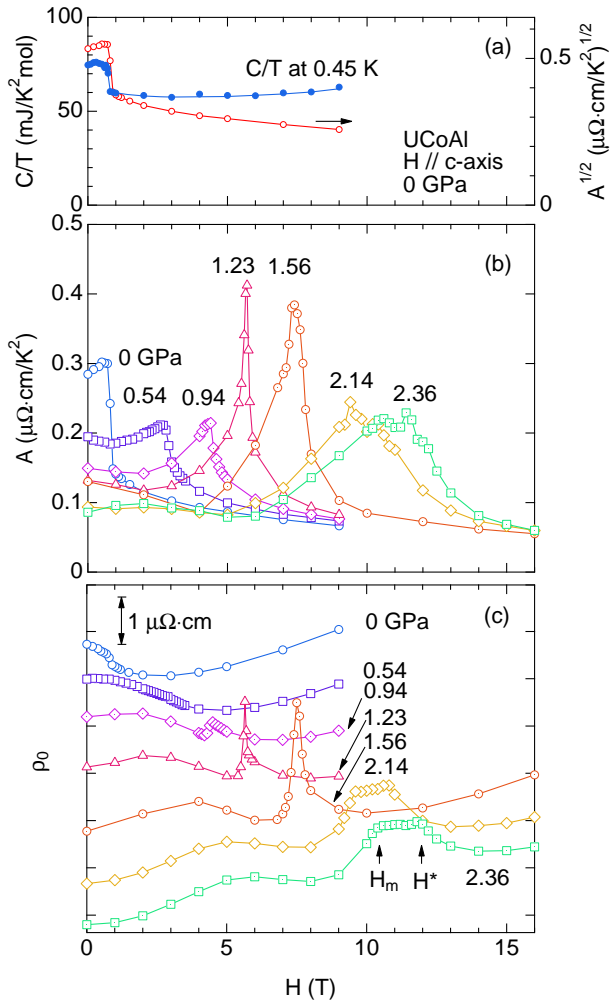}
\end{center}
\caption{(Color online) (a)Field dependence of the specific heat at $0.45\,{\rm K}$ for $H\parallel c$-axis in UCoAl. (b)Field dependence of the resistivity $A$ coefficient at various pressures and (c) corresponding residual resistivity. The data in panel (c) is vertically shifted for clarity. In panel (a), the field dependence of $A$ coefficient is plotted, as well, in the form of $\sqrt{A}$ vs $H$, assuming Kadowaki-Woods relation.}
\label{fig:UCoAl_A_coef}
\end{figure}

From the temperature dependence of the resistivity at constant field,
we determined the field dependence of the resistivity $A$ coefficient and the residual resistivity $\rho_0$
at different pressures, as shown in Fig.~\ref{fig:UCoAl_A_coef}(b)(c).
Here the resistivity is described by $\rho = \rho_0 + AT^2$.
Considering the Kadowaki-Woods relation ($A \propto \gamma^2$, $\gamma$: Sommerfeld coefficient) based on the existence of the strong local fluctuation,
the field dependence of the $A$ coefficient corresponds to that of the square of the effective mass $m^\ast$ ($A\propto \gamma^2 \propto {m^\ast}^2$).
For comparison, 
the field dependence of the $\gamma$-value at ambient pressure estimated from the measurements at $0.45\,{\rm K}$ is also shown 
in Fig.~\ref{fig:UCoAl_A_coef}(a).
The results are consistent with the previous results~\cite{Mat99}
The $\gamma$-value at zero field is $75\,{\rm mJ/K^2 mol}$, which is almost unchanged up to $H_{\rm m}$.
Further increasing field, the $\gamma$-value is abruptly reduced down to $60\,{\rm mJ/K^2 mol}$ and is almost constant up to $9\,{\rm T}$.
It should be noted that the slight upturn at high fields is due to the hyperfine contribution from Co and Al.
This is roughly consistent with the field dependence of $A$ coefficient at ambient pressure shown in Fig.~\ref{fig:UCoAl_A_coef}(a).
However, the normalization to the high field limit will lead to an enhancement of $A$ at zero field
in agreement with the enhancement of $\sqrt{A}$ with respect to $\gamma$ in ferromagnetic spin fluctuation theory.

With increasing pressure, the field dependence of the $A$ coefficient is drastically changed, as shown in Fig.~\ref{fig:UCoAl_A_coef}(b).
$H_{\rm m}$ increases linearly with field, and the $A$ coefficient at zero field is suppressed with pressure.
Instead, the peak structure at $H_{\rm m}$ becomes pronounced.
At $1.23\,{\rm GPa}$, the $A$ coefficient reveals a very sharp peak at $H_{\rm m}$, 
which is approximately three times larger than that at zero field.
Further increasing pressure, the peak value of $A$ coefficient is reduced
and the width of the peak is significantly increased.
Above $P_{\rm QCEP}\sim 1.5\,{\rm GPa}$, 
the $A$ coefficient exhibits a plateau.

\begin{figure}[tbh]
\begin{center}
\includegraphics[width=1 \hsize,clip]{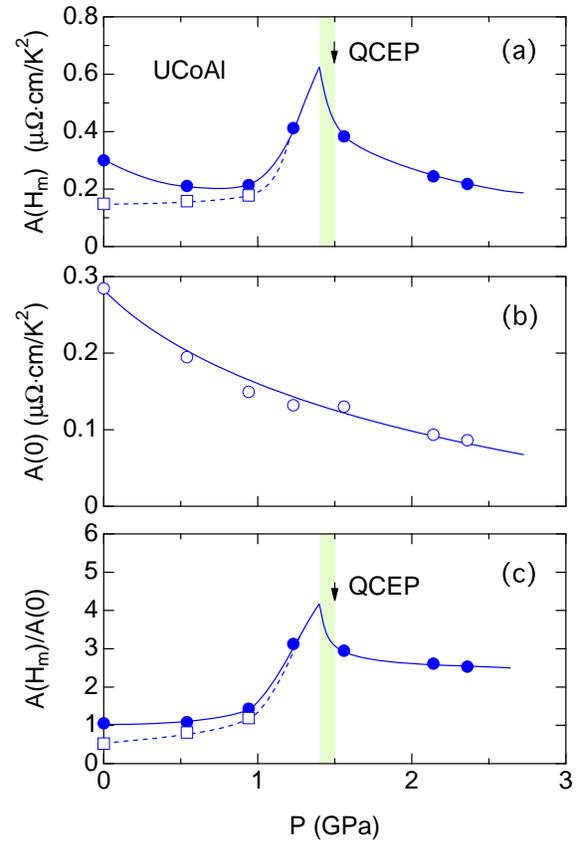}
\end{center}
\caption{(Color online) Pressure dependence of the $A$ coefficient (a) at $H_{\rm m}$ and (b) at zero field in UCoAl. (c)Pressure dependence of the ratio of $A(H_{\rm m})$ to $A(0)$. Closed symbols and open symbols in panel (a) and (c) correspond to the peak value and dropped value at $H_{\rm m}$, respectively. The lines are guides to the eyes.}
\label{fig:UCoAl_A_coef_Hm}
\end{figure}

In order to see the pressure evolution of the $A$ coefficient more clearly,
we plot the pressure dependence of the $A$ coefficient at $H_{\rm m}$ and at zero field,
as shown in Fig.~\ref{fig:UCoAl_A_coef_Hm}.
The $A$ coefficient at zero field, $A(0)$, monotonously decreases in Fig.~\ref{fig:UCoAl_A_coef_Hm}(b),
indicating that the pressure drives UCoAl away from the critical region.
On the other hand, the $A$ coefficient at $H_{\rm m}$, $A(H_{\rm m})$, shows a maximum around the QCEP.
If we take the ratio, $A(H_{\rm m})/A(0)$, the enhancement of $A(H_{\rm m})$ at QCEP is more significant, as shown in Fig.~\ref{fig:UCoAl_A_coef_Hm}(c).
It is interesting to note that the value of $A(H_{\rm m})/A(0)$ seems to remain constant above $P_{\rm QCEP}$.

The residual resistivity shows a plateau, as well (see Fig.~\ref{fig:UCoAl_A_coef}(c)).
The kink of the plateau at lower field, for example, $H_{\rm m}=10.5\,{\rm T}$ at $2.36\,{\rm GPa}$
corresponds to the continuation of $H_{\rm m}$, which
was detected by the magnetostriction (see Fig.~\ref{fig:UCoAl_Pdep_jump}(a)).
However, the kink of the plateau at higher field (ex. $H^\ast=12\,{\rm T}$ at $2.36\,{\rm GPa}$)
was detected only by the resistivity, but the magnetostriction down to $0.3\,{\rm K}$ shows no anomaly.
Interestingly, the residual resistivity exhibits a sharp peak around $P_{\rm QCEP}$.
This is in good agreement with the prediction of residual resistivity enhancement at the ferromagnetic singularity~\cite{Miy02}.
Furthermore, a broad anomaly is observed around $6\,{\rm T}$ at $2.36\,{\rm GPa}$,
which shifts to lower field with decreasing pressure.
These broad anomalies might be explained by the competitive phenomena 
between the scattering near $H_{\rm m}$ and the orbital effect of transverse magnetoresistance,
which is related to the value of $\omega_{\rm c}\tau$ where $\omega_{\rm c}$ ($=eH/m^\ast c$) and $\tau$ are
the cyclotron frequency and the scattering lifetime, respectively.

\begin{figure}[tbh]
\begin{center}
\includegraphics[width=1 \hsize,clip]{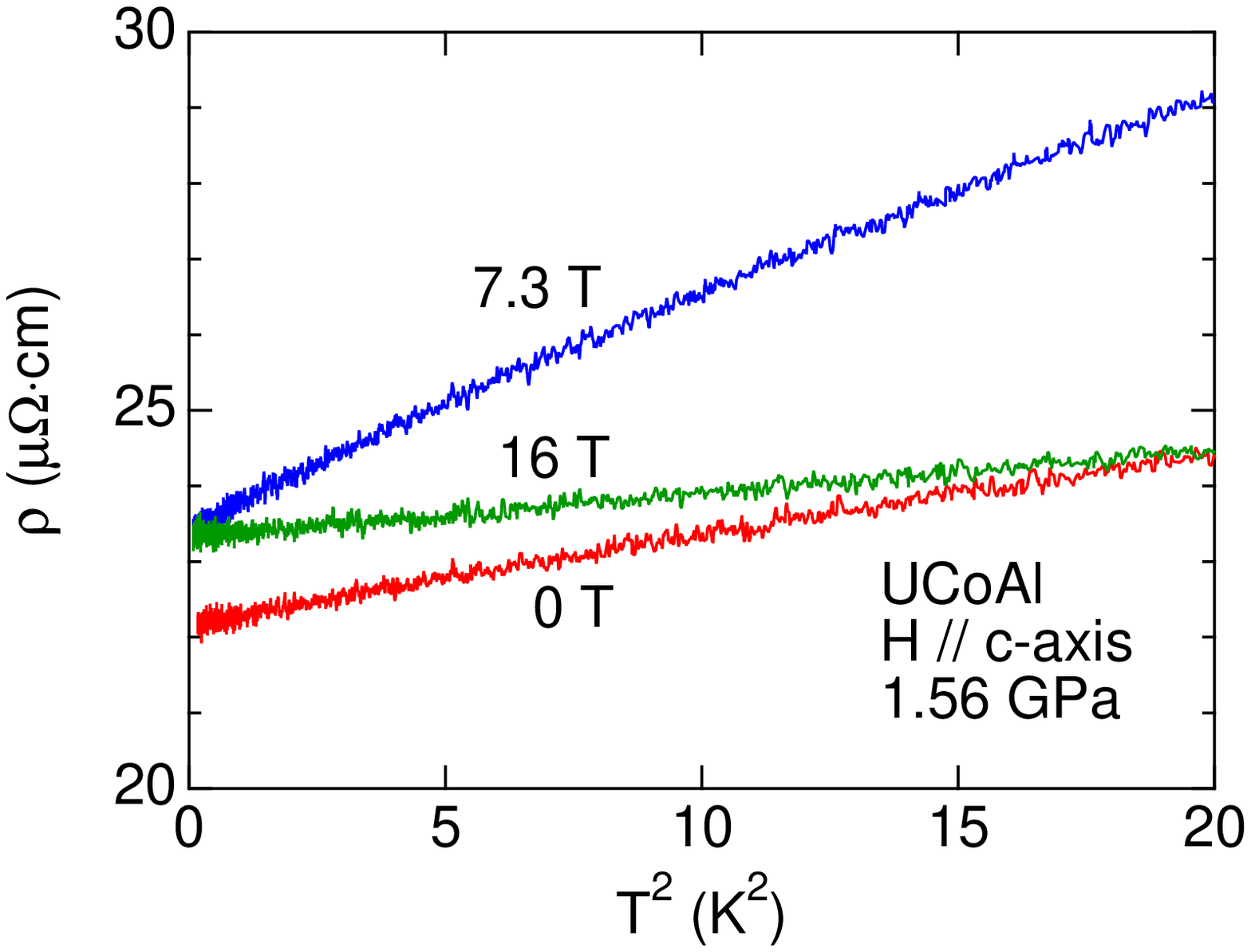}
\end{center}
\caption{(Color online) Temperature dependence of the resistivity at $0$, $7.3$ ($\sim H_{\rm m}$) and $16\,{\rm T}$ for $H\parallel c$-axis under pressure at $1.56\,{\rm GPa}$ in UCoAl.}
\label{fig:UCoAl_Tdep_resist}
\end{figure}

It should be noted that temperature range where the resistivity follows a $T^2$ relation
becomes narrower around QCEP, as shown in Fig.~\ref{fig:UCoAl_Tdep_resist}.
For example, at $1.56\,{\rm GPa}$, $T^2$ behavior was observed only up to $\sim 1\,{\rm K}$ at $7.3\,{\rm T}$.
An interesting point is to investigate the critical exponent of resistivity at the QCEP,
which is left for future studies of precise resistivity measurements.
%========================================================================================
%========================================================================================
%========================================================================================
%========================================================================================
%========================================================================================
%========================================================================================

\begin{figure}[tbh]
\begin{center}
\includegraphics[width=1 \hsize,clip]{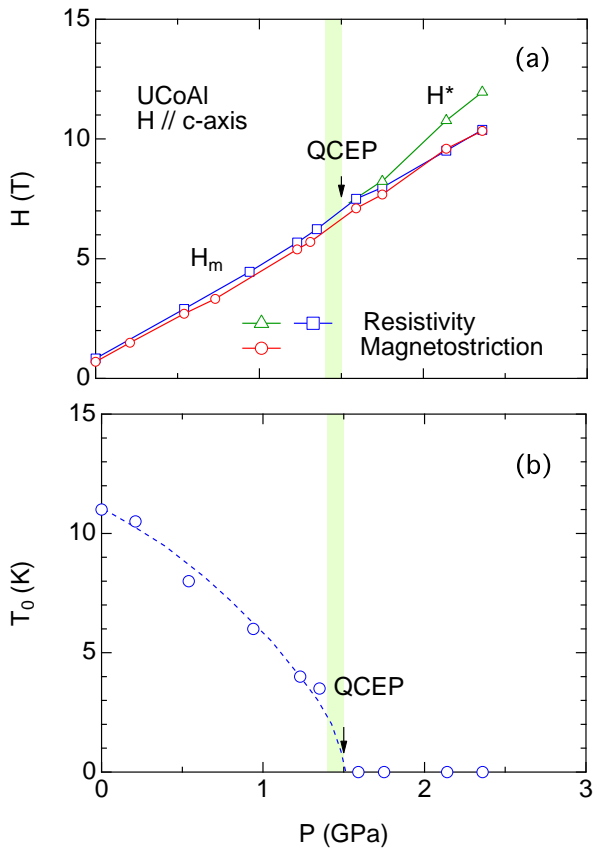}
\end{center}
\caption{(Color online) (a)Pressure dependence of $H_{\rm m}$ extrapolated to $0\,{\rm K}$ for $H\parallel c$-axis in UCoAl,
and (b)the pressure dependence of $T_{\rm 0}$ where the 1st order transition changes into the crossover.}
\label{fig:UCoAl_Hm_phase}
\end{figure}

Figure~\ref{fig:UCoAl_Hm_phase} shows the pressure dependence of $H_{\rm m}$ extrapolated to $0\,{\rm K}$
obtained by the magnetoresistance and magnetostriction measurements.
$H_{\rm m}$ is almost linearly increased with pressure up to $7\,{\rm T}$ at $P_{\rm QCEP}\sim 1.5\,{\rm GPa}$.
The value of $H_{\rm m}$ from the resistivity measurements almost coincides with that from the magnetostriction measurements,
although $H_{\rm m}$ of resistivity is slightly higher than that of magnetostriction,
which is most likely due to the slight pressure inhomogeneity in the pressure cell.
Above $P_{\rm QCEP}$, the magnetostriction shows the linear increase of $H_{\rm m}$ continuously,
while the magnetoresistance shows the split corresponding to the plateau of the residual resistivity.
The lower field anomaly is in good agreement with the results of magnetostriction.

Figure~\ref{fig:UCoAl_Hm_phase}(b) shows the pressure dependence of $T_0$
at which the 1st order transition of $H_{\rm m}$ terminates and changes into a crossover.
The value was evaluated by the field derivative of magnetoresistance, as shown in the inset of Fig.~\ref{fig:UCoAl_resist_ambient}(b).
$T_{\rm 0}$ decreases with pressure, and collapses around $P_{\rm QCEP}\sim 1.5\,{\rm GPa}$,
indicating that the 1st order transition at $H_{\rm m}$ terminates at $P_{\rm QCEP}$,
and a new phase appears above $P_{\rm QCEP}$.
%========================================================================================
%========================================================================================

\begin{figure}[tbh]
\begin{center}
\includegraphics[width=1 \hsize,clip]{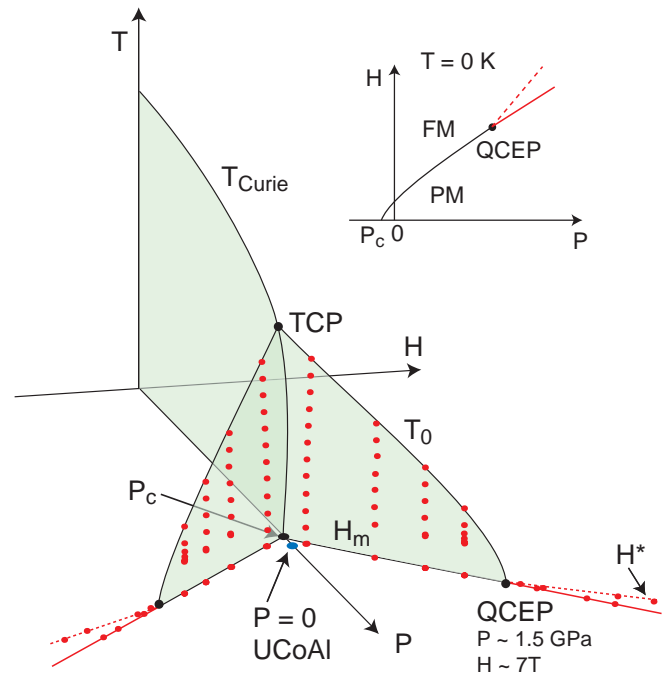}
\end{center}
\caption{(Color online) Semi-schematic temperature-pressure-field phase diagram. Red closed circles correspond to the experimental data points. The Curie temperature $T_{\rm Curie}$ is suppressed with pressure and bifurcates at the tricritical point (TCP), where the transition changes from 2nd order to 1st order, revealing the 1st order plane. The boundary between the 1st order and the crossover at finite temperature is called $T_0$.
UCoAl at ambient pressure is located just above the critical pressure $P_{\rm c}$ at which $T_{\rm Curie}$ collapses at $H=0$. Applying magnetic fields, the metamagnetic transition occurs at $H_{\rm m}$ crossing the 1st order plane. Increasing pressure, $H_{\rm m}$ increases, while $T_0$ decreases and terminate at the quantum critical endpoint (QCEP). Above $P_{\rm QCEP}$ $H_{\rm m}$ continuously increases. From QCEP, new anomaly appears, as indicated by the dashed line. It is noted that the data points at ``negative'' fields are depicted as a mirror of those at ``positive'' field for the comparison with a generic phase diagram near FM instabilities~\protect\cite{Bel05}.}
\label{fig:UCoAl_TPH}
\end{figure}

As a summary of pressure experiments,
we schematically show in Fig.~\ref{fig:UCoAl_TPH} the temperature--pressure--field phase diagram of UCoAl,
together with the field--pressure phase diagram at $0\,{\rm K}$.
The critical point where $T_{\rm Curie}$ is suppressed to $0\,{\rm K}$ is at a negative pressure ($P_{\rm c}\sim -0.2\,{\rm GPa}$) in UCoAl.
At the tricritical point (TCP), $T_{\rm Curie}$ bifurcates and the 1st order plane appears.
When the magnetic field is applied at ambient pressure, 
UCoAl crosses the 1st order plane, which corresponds to the metamagnetic transition
from paramagnetic ground state to the field-induced ferromagnetic state.
The temperature, which is located on the 1st order plane at finite temperature, corresponds to $T_0$.
Increasing pressure, $H_{\rm m}$ increases linearly and meets with QCEP
where $T_0$ is suppressed to $0\,{\rm K}$.
At QCEP, the effective mass shows the acute enhancement.
At higher pressure $P > P_{\rm QCEP}$, 
$H_{\rm m}$ increases continuously as a crossover,
which can be observed both by resistivity and by magnetostriction.
From the QCEP, a new transition or crossover line at $H^\ast$ which was detected in $\rho_0$ appears and deviates from
the original $H_{\rm m}$ line.
%========================================================================================
%========================================================================================

%========================================================================================
%========================================================================================
\section{Discussion}
A general treatment~\cite{Min11_UGe2} implies that the ferromagnetic phase diagram with a strong decrease of $T_{\rm Curie}$ near $P_{\rm c}$ 
will be extended in magnetic field by a first order line $T_0$ which will terminate at a pressure higher than $P_{\rm c}$.
In a conventional approach, the paramagnetic--ferromagnetic transition at $H_{\rm m}$
has been described with special features in the density of states such as a double peak structure or van Hove singularity.~\cite{San03,Bin04,Shi64}
It has also been treated in spin fluctuation theory assuming Fermi surface invariance through $H_{\rm m}$ 
or quite generally by considering excitations at the Fermi level which introduce non-analytic corrections 
in the Landau expression of the free energy~\cite{Bel05,Mil02,Yam93_meta}
It was, however, recently stressed that a strong modification of the Fermi surface can occur at $H_{\rm m}$.
A recent calculation assuming that a Lifshitz transition may occur at $H_{\rm m}$ shows that
the ferromagnetic wing structure on the phase diagram will be strongly affected in this case.~\cite{Yam07_Lifshitz}

% comparison with UGe2
Here we compare the present results of UCoAl with those recently obtained for the other Ising ferromagnet UGe$_2$.
At ambient pressure, UGe$_2$ has a high Curie temperature $T_{\rm Curie}\sim 52\,{\rm K}$
associated with a large ferromagnetic ordered moment $M_0 \sim 1.5\,\mu_{\rm B}$.
The tricritical point (TCP) is located at $T_{\rm TCP}\sim 24\,{\rm K}$ and $P_{\rm TCP}\sim 1.42\,{\rm GPa}$.
The ferromagnetism disappears at $P_{\rm c}\sim 1.5\,{\rm GPa}$ at zero field via
a sharp drop of the magnetization $\Delta M_0 \sim 0.9\,\mu_{\rm B}$ with a 1st order nature.~\cite{Pfl02}
Due to the large value of $\Delta M_0$ in UGe$_2$, the QCEP is achieved 
at  high pressure ($P_{\rm QCEP}\sim 3.5\,{\rm GPa}\sim 2 P_{\rm c}$) and high field ($H_{\rm QCEP}\sim 18\,{\rm T}$).

\begin{figure}[tbh]
\begin{center}
\includegraphics[width=1 \hsize,clip]{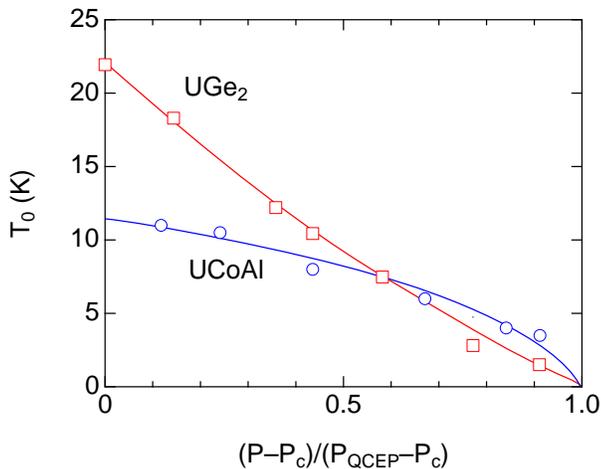}
\end{center}
\caption{(Color online) Scaled pressure dependence of $T_0$ of UCoAl and UGe$_2$. In UCoAl, the critical pressure $P_{\rm c}$ and the pressure at QCEP are $-0.2\,{\rm GPa}$ and $1.5\,{\rm GPa}$, respectively. In UGe$_2$, $P_{\rm c}$ and $P_{\rm QCEP}$ are $1.5\,{\rm GPa}$ are $3.6\,{\rm GPa}$, respectively. The data of UGe$_2$ are cited from ref.\protect\citen{Kot11}.}
\label{fig:UCoAl_UGe2_T0}
\end{figure}

Figure~\ref{fig:UCoAl_UGe2_T0} shows the pressure dependence of $T_0$ as a function of the scaled pressure, $(P-P_{\rm c})/(P_{\rm QCEP}-P_{\rm c})$
in UCoAl and UGe$_2$~\cite{Kot11}.
In UGe$_2$, $T_0$ has an upward (concave) curvature, while in UCoAl $T_0$ shows a downward (convex) curvature.
This difference might correspond to a difference of Fermi surface dimensionality between the two compounds.
In the ferromagnetic state of UGe$_2$, both dHvA experiments~\cite{Sat92b,Ter01,Set02,Hag02,Ter02,Set03} and band structure calculation~\cite{Yam93,Shi01_UGe2} show a quasi-two dimensional Fermi surface at least for a main dHvA branch.
Such lower dimensionality can explain the upward curvature of $T_0$.~\cite{Mil02}
Up to now, there are no reports concerning Fermi surface topology on UCoAl,
although band structure calculations have been reported.~\cite{Eri89,Bet00}
However, three dimensionality is expected from the crystal structure.
It should be noted that the crystal structure without inversion symmetry in UCoAl may also affect the pressure response of $T_0$.
%========================================================================================
%========================================================================================

\begin{figure}[tbh]
\begin{center}
\includegraphics[width=1 \hsize,clip]{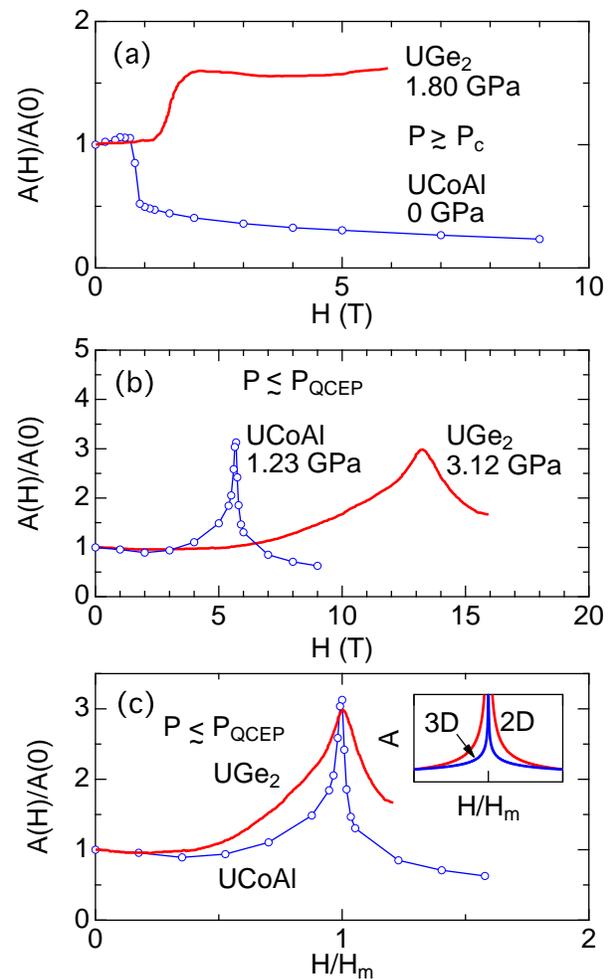}
\end{center}
\caption{(Color online) Field dependence of the $A$ coefficient (a) just above $P_{\rm c}$ and (b) just below $P_{\rm QCEP}$ in UCoAl and UGe$_2$.
$A$ coefficient is normalized with the value of zero field. (c)Field is normalized with $H_{\rm m}$ from panel (b). The inset of panel (c) shows the prediction from the spin fluctuation theory for two and three dimensional cases. 
The data of UGe$_2$ are cited from ref.\protect\citen{Kot11}.}
\label{fig:UCoAl_UGe2_A}
\end{figure}

%%%%%%% A coefficient: comparison 
The difference between the two compounds is confirmed by the field dependence of the resistivity $A$ coefficient,
as shown in Fig.~\ref{fig:UCoAl_UGe2_A}.
In UGe$_2$, the $A$ coefficient shows a step-like increase at $H_{\rm m}$ at pressures just above $P_{\rm c}$ 
($P=1.8\,{\rm GPa} > P_{\rm c}$),~\cite{Kot11} as shown in Fig.\ref{fig:UCoAl_UGe2_A}(a),
while the $A$ coefficient near $P_{\rm c}$ in UCoAl shows a step-like decrease at $H_{\rm m}$.
However, near $P_{\rm QCEP}$, the field dependence of the $A$ coefficient for both UCoAl and UGe$_2$ shows a similar peak structure
at $H_{\rm m}$, where the enhanced values for both compounds are three times larger than those at zero field,
although the peak of UCoAl is much shaper than that of UGe$_2$.
This observation is consistent with the effect of lower dimensionality in UGe$_2$.
It is indeed predicted as 
$A\sim (|H-H_{\rm QCEP}|/H_{\rm QCEP} )^{-1/3}$ for $d=3$ (3D system) and
$A\sim (|H-H_{\rm QCEP}|/H_{\rm QCEP} )^{-2/3}$ for $d=2$ (2D system),~\cite{Mil02} as shown in the inset of Fig.~\ref{fig:UCoAl_UGe2_A}(c)
but for both cases, the data cannot be fully fitted with a spin fluctuation approach.
A possible reason might be unusual critical exponent around $H_{\rm QCEP}$.
The present results of $A$ coefficient are obtained, assuming that the resistivity follows $T^2$ at low temperatures just above $0.1\,{\rm K}$.
More precise measurements at lower temperatures is left for the future study.
That will allow a definitive comparison with a spin fluctuation approach.
%========================================================================================
%========================================================================================

The difficult question is possible changes of the Fermi surface at $H_{\rm m}$.
In UGe$_2$, the drastic change of the Fermi surface between the paramagnetic state and the ferromagnetic state (FM1) 
was directly observed by de Haas-van Alphen (dHvA) experiments.~\cite{Ter01,Set02,Hag02,Ter02,Set03}
Nevertheless, it is still an open question whether the Fermi surface changes 
near the QCEP because of the large values of $P_{\rm QCEP}$ and $H_{\rm QCEP}$.
The change of Fermi surface associated with a Lifshitz transition near $P_{\rm QCEP}$
might be a possible origin for the deviation of $T_{\rm 0}$ and $A$ from the spin fluctuation theory.

Up to now, no clear quantum oscillation experiments has been reported for UCoAl.
However, the Hall coefficient and thermopower coefficient at ambient pressure reveal a drastic change through $H_{\rm m}$,~\cite{Mat00}
also suggesting a change of Fermi surface.
An interesting point in UCoAl is that the measurements can be easily achieved far above $P_{\rm QCEP}$.

%%%%% Imada and Lifshits transition
A new feature in the present study is that further increasing pressure above QCEP, 
the transition line continues as a crossover line from the 1st order metamagnetic transition
and a new high field phase appears.
Theoretically, a new mechanism caused by the topological change of Fermi surface related to the Lifshitz-type transition
is proposed at QCEP, where the unusual behavior ($d\chi^{-1}/dM \to \infty$, $\chi$: susceptibility, $M$: magnetization) is predicted.~\cite{Yam07_Lifshitz,Kim04}

%%%% Interpretation of plateau
Another interesting result is that the field dependence of $\rho_0$ appears to have a plateau for $P > P_{\rm QCEP}$
associated with a plateau in the field dependence of $A$.
For example at $2.36\,{\rm GPa}$,
the plateau extends from $10$ to $12\,{\rm T}$.
It should be noted that the present experimental setup means transverse magnetoresistance,
which can be affected by the orbital effect of the cyclotron motion, namely $\omega_{\rm c}\tau$ 
($\omega_{\rm c}=eH/m_{\rm c}^\ast c$).
If the orbit is associated with van Hove singularity and a Lifshitz transition,
a feedback can occur on the variation of the cyclotron effective mass $m_{\rm c}^\ast$.

In general, a field sweep on a system with a sharp singularity in the density of states 
can lead to a large variety of phenomena: 
collapse or change of interactions (balance between ferromagnetic and antiferrmagnetic channel),
field dependence of local fluctuations (Kondo effect and valence fluctuations)~\cite{Flo06_review,Wat09}
and
field evolution of Fermi surfaces (Lifshitz transition and/or Pauli depairing of small Fermi surface sheets).~\cite{Shi09,Alt11,Mal11}
A recent appealing possibility will be the occurrence of nematic phase as suggested in Sr$_3$Ru$_2$O$_7$~\cite{Bor07,Fra10}

A simple way to understand the plateau is to assume two different contributions;
one is located at $H_{\rm m}$ and the other is located at $H^\ast$.
For example, through the studies of metamagnetic phenomena in CeRu$_2$Si$_2$ family,
a plateau of the field dependence of effective mass is detected 
in Ce(Ru$_{0.92}$Rh$_{0.08}$)$_2$Si$_2$ 
between $H_{\rm c}=3\,{\rm T}$ and $H_{\rm m}=5.5\,{\rm T}$.~\cite{Sek92}
The details will be published elsewhere.
In the pure system CeRu$_2$Si$_2$, it is known that two singularities attributed to antiferromagnetic fluctuations and the ferromagnetic fluctuations
occur at the same field $H_{\rm m}\sim 7\,{\rm T}$~\cite{Aok11_CeRu2Si2}, as detected by neutron scattering experiments.~\cite{Sat04}
However, in Ce(Ru$_{0.92}$Rh$_{0.08}$)$_2$Si$_2$, two singularities might be separated.
Although the microscopic experimental evidence is not yet obtained,
the simple image is that the nature of antiferromagnetic order is strongly modified by Rh-doping 
going from the transverse (La-doping) to longitudinal antiferromagnetic mode at zero field.
However, the fact that the magnetic field favors the transverse mode in antiferromagnetic correlations
leads to a switch in the paramagnetic phase with only strong antiferromagnetic correlations at $H_{\rm c}$;
the subsequent crossing through a regime dominated by ferromagnetic fluctuations occurs at $H_{\rm m}$ which is higher than $H_{\rm c}$.
In UCoAl, similar two contributions may give rise to the plateau of effective mass.

%%%% comparison with Sr-327
The plateau observed in UCoAl above $P_{\rm QCEP}$ has some similarities 
with the results of Sr$_3$Ru$_2$O$_7$, which are interpreted as a signature of a nematic fluid 
which is characterized as a translationally invariant metallic phase with a spontaneously generated spatial anisotropy.~\cite{Fra10}
This proposal supports the idea that strongly correlated electrons can self-organize in quite different fashions
and the metamagnetism here is not the only possibility.
The possibility of a nematic phase in Sr$_3$Ru$_2$O$_7$ has been mainly given by the field-angle tuned magnetoresistance
and the anisotropy.~\cite{Bor07}
Recent pressure experiments have clearly shown that the uniform magnetization density is not the order parameter near QCEP
but clear signatures of nematic phase were not observed.~\cite{Wu11}
It is worthwhile to remark that UCoAl is a Ising 3D system with respect to the Fermi surface properties,
while Sr$_3$Ru$_2$O$_7$ is a Heisenberg type with 2D Fermi surfaces.
As shown in Figs.~\ref{fig:UCoAl_sus_mag} and \ref{fig:UCoAl_MS_angdep}, 
UCoAl has strong Ising nature where $H_{\rm m}$ strongly increases with a $1/\cos\theta$ dependence,
while $H_{\rm m}$ in Sr$_3$Ru$_2$O$_7$ shows 
the moderate increase from $5\,{\rm T}$ for $H\parallel ab$-plane to $\sim 7.8\,{\rm T}$ for $H\parallel c$-axis.
Basically, in UCoAl at least at low pressures, the key ingredient is the component of the magnetization along the $c$-axis.
There is no relation between field-angle and pressure for singularities.
An interesting question is whether the Ising type behavior is changed into the quasi-Heisenberg type at high pressure above QCEP, 
together with the topological change of Fermi surface. 
A conservative view is to assume that the Ising character is preserved through QCEP
and thus the plateau observed above $P_{\rm QCEP}$ cannot be associated with a nematic phase.
Key experiments will be magnetization, Hall effect, and thermopower measurements under pressure at high fields.
If UCoAl will be a weak ferromagnet, as it is the case for URhGe~\cite{Lev07,Miy09,Aok11_ICHE} and UCoGe~\cite{Aok09_UCoGe,Aok11_JPSJ_review}
one can expect that transverse field perpendicular to $M_0$ will have strong effect on the ferromagnetic instability.
Here excellent agreement is found with the view that tilting the field-angle only modifies the Zeeman energy.

%-----------------------------------------------------------------------
\section{Summary}
We grew single crystals of UCoAl and performed resistivity and magnetostriction measurements under pressure up to $2.4\,{\rm GPa}$ and at high fields up to $16\,{\rm T}$.
The metamagnetic transition at $H_{\rm m}$ changes from the 1st order at low temperature to the crossover at high temperature.
The critical temperature $T_0$ is determined by the field sweep of resistivity and magnetostriction measurements.
With increasing pressure, $H_{\rm m}$ monotonously increases, while $T_0$ decreases and is suppressed at the quantum critical endpoint (QCEP).
The field dependence of the effective mass detected by the resistivity $A$ coefficient
reveals the acute enhancement at QCEP at $H_{\rm m}$.
Further increasing pressure, $H_{\rm m}$ increases continuously as the crossover,
which was detected both by resistivity and magnetostriction.
The resistivity measurements exhibit another new anomaly at higher field $H^\ast$ than $H_{\rm m}$ 
at pressures above $P_{\rm QCEP}$.
Our experiments show that UCoAl will be a key example of field-induced state built from ferromagnetic fluctuations.
Its rather low values of $P_{\rm QCEP}\sim 1.5\,{\rm GPa}$ and $H_{\rm QCEP}\sim 7\,{\rm T}$ will allow one to
perform a large variety microscopic and macroscopic experiments,
which can provide definitive conclusions on the properties of QCEP and in the plateau regime detected at $P > P_{\rm QCEP}$.

%%%%%%%%%%%%%%%%%%%%%%%%%%%%%%%%%%%%%%%%%%%%%%%%%%%%%%%%%%%%%%%%%%%%%%%%%%%%%%%%%

%%%%%%%%%%%%%%%%%%%%%%%%%%%%%%%%%%%%%%%%%%%%%%%%%%%%%%%%%%%%%%%%%%%%%%%%%%%%%%%%%
\section*{Acknowledgements}
%%%%%%%%%%%%%%%%%%%%%%%%%%%%%%%%%%%%%%%%%%%%%%%%%%%%%%%%%%%%%%%%%%%%%%%%%%%%%%%%%
We thank H. Harima, L. Malone and A. P. Mackenzie for useful discussions.
This work was supported by ERC starting grant (NewHeavyFermion), French ANR project (CORMAT, SINUS, DELICE).

%\bibliographystyle{myjpsj}
%\bibliography{bibbase}

\begin{thebibliography}{10}

\bibitem{Flo06_review}
J.~Flouquet: {\em Progress in Low Temperature Physics}, ed. W.~P. Halperin
  (Elsevier, Amsterdam, 2006) p.~139.

\bibitem{Flo10}
J.~Flouquet, D.~Aoki, W.~Knafo, G.~Knebel, T.~D. Matsuda, S.~Raymond,
  C.~Proust, C.~Paulsen and P.~Haen: J. Low Temp. Phys. {\bf 161} (2010) 83.

\bibitem{Fis91}
R.~A. Fisher, C.~Marcenat, N.~E. Phillips, P.~Haen, F.~Lapierre, P.~Lejay,
  J.~Flouquet and J.~Voiron: J. Low Temp. Phys. {\bf 84}~({1-2}) (1991) 49.

\bibitem{Aok11_CeRu2Si2}
D.~Aoki, C.~Paulsen, T.~D. Matsuda, L.~Malone, G.~Knebel, P.~Haen, P.~Lejay,
  R.~Settai, Y.~\={O}nuki and J.~Flouquet: J. Phys. Soc. Jpn. {\bf 80} (2011)
  053702.

\bibitem{Tau10}
V.~Taufour, D.~Aoki, G.~Knebel and J.~Flouquet: Phys. Rev. Lett. {\bf 105}~(21)
  (2010) 217201.

\bibitem{Kot11}
H.~Kotegawa, V.~Taufour, D.~Aoki, G.~Knebel and J.~Flouquet: to be published in
  J. Phys. Soc. Jpn.

\bibitem{Uhl04}
M.~Uhlarz, C.~Pfleiderer and S.~M. Hayden: Phys. Rev. Lett. {\bf 93}~(25)
  (2004) 256404.

\bibitem{Gri01}
S.~A. Grigera, R.~S. Perry, A.~J. Schofield, M.~Chiao, S.~R. Julian, G.~G.
  Lonzarich, S.~I. Ikeda, Y.~Maeno, A.~J. Millis and A.~P. Mackenzie: Science
  {\bf 294} (2001) 329.

\bibitem{Sax00}
S.~S. Saxena, P.~Agarwal, K.~Ahilan, F.~M. Grosche, R.~K.~W. Haselwimmer, M.~J.
  Steiner, E.~Pugh, I.~R. Walker, S.~R. Julian, P.~Monthoux, G.~G. Lonzarich,
  A.~Huxley, I.~Sheikin, D.~Braithwaite and J.~Flouquet: Nature {\bf 406}
  (2000) 587.

\bibitem{She01}
I.~Sheikin, A.~Huxley, D.~Braithwaite, J.~P. Brison, S.~Watanabe, K.~Miyake and
  J.~Flouquet: Phys. Rev. B {\bf 64} (2001) 220503.

\bibitem{And85}
A.~V. Andreev, R.~Z. Levitin, Y.~F. Popov and R.~Y. Yumaguzhin: Sov. Phys.
  Solid State {\bf 27} (1985) 1145.

\bibitem{And07}
A.~V. Andreev, K.~Koyama, N.~V. Mushnikov, V.~Sechovsk{\'y}, Y.~Shiokawa,
  I.~Satoh and K.~Watanabe: J. Alloys Comp. {\bf 441} (2007) 33.

\bibitem{Ish03}
Y.~Ishii, M.~Kosaka, Y.~Uwatoko, A.~V. Andreev and V.~Sechovsk{\'y}: Physica B
  {\bf 334} (2003) 160.

\bibitem{Mus99}
N.~V. Mushnikov, T.~Goto, K.~Kamishima, H.~Yamada, A.~V. Andreev, Y.~Shiokawa,
  A.~Iwao and V.~Sechovsk{\'y}: Phys. Rev. B {\bf 90} (1999) 6877.

\bibitem{Koy07}
K.~{Koyama-Nakazawa}, M.~Koeda, M.~Hedo and Y.~Uwatoko: Rev. Sci. Instr. {\bf
  78} (2007) 066109.

\bibitem{Yok07}
K.~Yokogawa, K.~MURATA, H.~Yoshino and S.~Aoyama: Jpn. J. Appl. Phys. {\bf 46}
  (2007) 3636.

\bibitem{Hav97}
L.~Havela, A.~V. Andreev, V.~Sechovsk{\'y}, I.~K. Kozlovskaya, K.~Proke{\v{s}},
  P.~Javorsk{\'y}, M.~I. Bartashevich, T.~Goto and K.~Kamishima: Physica B {\bf
  230-232} (1997) 98.

\bibitem{Mat99}
T.~D. Matsuda, Y.~Aoki, H.~Sugawara, H.~Sato, A.~V. Andreev and V.~Sechovsky:
  J. Phys. Soc. Jpn. {\bf 68} (1999) 3922.

\bibitem{Mus01}
N.~V. Mushnikov, A.~V. Andreev, T.~Goto and V.~Sechovsk{\'y}: Philos. Mag. B
  {\bf 81} (2001) 569.

\bibitem{Aok01}
D.~Aoki, A.~Huxley, E.~Ressouche, D.~Braithwaite, J.~Flouquet, J.-P. Brison,
  E.~Lhotel and C.~Paulsen: Nature {\bf 413} (2001) 613.

\bibitem{Lev07}
F.~L\'{e}vy, I.~Sheikin and A.~Huxley: Nature Physics {\bf 3} (2007) 460.

\bibitem{Miy09}
A.~Miyake, D.~Aoki and J.~Flouquet: J. Phys. Soc. Jpn. {\bf 78} (2009) 063703.

\bibitem{Aok11_ICHE}
D.~Aoki, T.~D. Matsuda, F.~Hardy, C.~Meingast, V.~Taufour, E.~Hassinger,
  I.~Sheikin, C.~Paulsen, G.~Knebel, H.~Kotegawa and J.~Flouquet: J. Phys. Soc.
  Jpn. {\bf 80} (2011) SA008.

\bibitem{Huy07}
N.~T. Huy, A.~Gasparini, D.~E. {de Nijs}, Y.~Huang, J.~C.~P. Klaasse,
  T.~Gortenmulder, A.~{de Visser}, A.~Hamann, T.~{G\"{o}rlach} and
  H.~v.~{L\"{o}hneysen}: Phys. Rev. Lett. {\bf 99} (2007) 067006.

\bibitem{Aok09_UCoGe}
D.~Aoki, T.~D. Matsuda, V.~Taufour, E.~Hassinger, G.~Knebel and J.~Flouquet: J.
  Phys. Soc. Jpn. {\bf 78} (2009) 113709.

\bibitem{Iha10}
Y.~Ihara, T.~Hattori, K.~Ishida, Y.~Nakai, E.~Osaki, K.~Deguchi, N.~K. Sato and
  I.~Satoh: Phys. Rev. Lett. {\bf 105} (2010) 206403.

\bibitem{Gri03}
S.~A. Grigera, R.~A. Borzi, A.~P. Mackenzie, S.~R. Julian, R.~S. Perry and
  Y.~Maeno: Phys. Rev. B {\bf 67} (2003) 214427.

\bibitem{Hon00}
F.~Honda, T.~Kagayama, G.~Oomi, L.~Havela, V.~Sechovsk{\'y} and A.~Andreev:
  Physica B {\bf 284-288} (2000) 1299.

\bibitem{Miy02}
K.~Miyake and O.~Narikiyo: J. Phys. Soc. Jpn. {\bf 71}~(3) (2002) 867.

\bibitem{Bel05}
D.~Belitz, T.~R. Kirkpatrick and {J\"{o}rg Rollb\"{u}hler}: Phys. Rev. Lett.
  {\bf 94}~(24) (2005) 247205.

\bibitem{Min11_UGe2} V. Mineev: C. R. Physique {\bf 12}, 567 (2011).

\bibitem{San03}
K.~G. Sandeman, G.~G. Lonzarich and A.~J. Schofield: Phys. Rev. Lett. {\bf
  90}~(16) (2003) 167005.

\bibitem{Bin04}
B.~Binz and M.~Sigrist: Europhys. Lett. {\bf 65} (2004) 816.

\bibitem{Shi64}
M.~Shimizu: Proc. Phys. Soc. {\bf 84} (1964) 397.

\bibitem{Mil02}
A.~J. Millis, A.~J. Schofield, G.~G. Lonzarich and S.~A. Grigera: Phys. Rev.
  Lett. {\bf 88} (2002) 217204.

\bibitem{Yam93_meta}
H.~Yamada: Phys. Rev. B {\bf 47}~(17) (1993) 11211.

\bibitem{Yam07_Lifshitz}
Y.~Yamaji, T.~Misawa and M.~Imada: J. Phys. Soc. Jpn. {\bf 76} (2007) 063702.

\bibitem{Pfl02}
C.~Pfleiderer and A.~D. Huxley: Phys. Rev. Lett. {\bf 89} (2002) 147005.

\bibitem{Sat92b}
K.~Satoh, S.~W. Yun, I.~Umehara, Y.~\={O}nuki, S.~Uji, T.~Shimizu and H.~Aoki:
  J. Phys. Soc. Jpn. {\bf 61} (1992) 1827.

\bibitem{Ter01}
T.~Terashima, T.~Matsumoto, C.~Terakura, S.~Uji, N.~Kimura, M.~Endo,
  T.~Komatsubara and H.~Aoki: Phys. Rev. Lett. {\bf 87}~(16) (2001) 166401.

\bibitem{Set02}
R.~Settai, M.~Nakashima, S.~Araki, Y.~Haga, T.~C. Kobayashi, N.~Tateiwa,
  H.~Yamagami and Y.~\={O}nuki: J. Phys.: Condens. Matter {\bf 14} (2002) L29.

\bibitem{Hag02}
Y.~Haga, M.~Nakashima, R.~Settai, S.~Ikeda, T.~Okubo, S.~Araki, T.~C. K.~N.
  Tateiwa and Y.~\={O}nuki: J. Phys.: Condens. Matter {\bf 14} (2002) L125.

\bibitem{Ter02}
T.~Terashima, T.~Matsumoto, C.~Terakura, S.~Uji, N.~Kimura, M.~Endo,
  T.~Komatsubara, H.~Aoki and K.~Maezawa: Phys. Rev. B {\bf 65} (2002) 174501.

\bibitem{Set03}
R.~Settai, M.~Nakashima, H.~Shishido, Y.~Haga, H.~Yamagami and Y.~\={O}nuki:
  Acta Physica Polonica B {\bf 34} (2003) 725.

\bibitem{Yam93}
H.~Yamagami and A.~Hasegawa: Physica B {\bf 186-188} (1993) 182.

\bibitem{Shi01_UGe2}
A.~B. Shick and W.~E. Pickett: Phys. Rev. Lett. {\bf 86}~(2) (2001) 300.

\bibitem{Eri89}
O.~Eriksson, B.~Johansson and M.~S.~S. Brooks: J. Phys.: Condens. Matter {\bf
  1} (1989) 4005.

\bibitem{Bet00}
K.~Betsuyaku and H.~Harima: Physica B {\bf 281-282} (2000) 778.

\bibitem{Mat00}
T.~D. Matsuda, H.~Sugawara, Y.~Aoki, H.~Sato, A.~V. Andreev, Y.~Shiokawa,
  V.~Sechovsky and L.~Havela: Phys. Rev. B {\bf 62} (2000) 13852.

\bibitem{Kim04}
N.~Kimura, M.~Endo, T.~Isshiki, S.~Minagawa, A.~Ochiai, H.~Aoki, T.~Terashima,
  S.~Uji, T.~Matsumoto and G.~G. Lonzarich: Phys. Rev. Lett. {\bf 92}~(19)
  (2004) 197002.

\bibitem{Wat09}
S.~Watanabe, A.~Tsuruta, K.~Miyake and J.~Flouquet: J. Phys. Soc. Jpn. {\bf
  78}~(10) (2009) 104706.

\bibitem{Shi09}
H.~Shishido, K.~Hashimoto, T.~Shibauchi, T.~Sasaki, H.~Oizumi, N.~Kobayashi,
  T.~Takamasu, K.~Takehana, Y.~Imanaka, T.~D. Matsuda, Y.~Haga, Y.~Onuki and
  Y.~Matsuda: Phys. Rev. Lett. {\bf 102}~(15) (2009) 156403.

\bibitem{Alt11}
M.~M. Altarawneh, N.~Harrison, S.~E. Sebastian, L.~Balicas, P.~H. Tobash, J.~D.
  Thompson, F.~Ronning and E.~D. Bauer: Phys. Rev. Lett. {\bf 106}~(14) (2011)
  146403.

\bibitem{Mal11}
L.~Malone, T.~D. Matusda, A.~Antunes, G.~Knebel, V.~Taufour, D.~Aoki,
  K.~Behnia, C.~Proust and J.~Flouquet: Phys. Rev. B {\bf 83} (2011) 245117.

\bibitem{Bor07}
R.~A. Borzi, S.~A. Grigera, J.~Farrell, R.~S. Perry, S.~J.~S. Lister, S.~L.
  Lee, D.~A. Tennant, Y.~Maeno and A.~P. Mackenzie: Science {\bf 315} (2007)
  214.

\bibitem{Fra10}
E.~Fradkin, S.~A. Kivelson, M.~J. Lawler, J.~P. Eisenstein and A.~P. Mackenzie:
  Annual Rev. Cond. Mat. Phys. {\bf 1} (2010) 153.

\bibitem{Sek92}
C.~Sekine, T.~Sakakibara, H.~Amitsuka and Y.~M. Goto: J. Phys. Soc. Jpn. {\bf
  61}~(12) (1992) 4536.

\bibitem{Sat04}
M.~Sato, Y.~Koike, S.~Katano, N.~Metoki, H.~Kadowaki and S.~Kawarazaki: J.
  Phys. Soc. Jpn. {\bf 73}~(12) (2004) 3418.

\bibitem{Wu11}
W.~Wu, A.~McCollam, S.~A. Grigera, R.~S. Perry, A.~P. Mackenzie and S.~R.
  Julian: Phys. Rev. B {\bf 83}~(4) (2011) 045106.

\bibitem{Aok11_JPSJ_review}
D.~Aoki and J.~Flouquet: to be published in J. Phys. Soc. Jpn.

\end{thebibliography}

\end{document}